\documentclass[aps,twocolumn,pra,prabib,
showpacs,preprintnumbers,amsmath,amssymb,floatfix]{revtex4}

\usepackage{graphicx}
\usepackage{dcolumn}
\usepackage{bm}

\newcommand{\bfr}{{\bf r}}

\newcommand{\gtwo}{g^{(2)}}

\begin{document}

\title{A proposal for measuring correlation functions in interacting gases} %

\author{C. Lobo}
\email{clobo@science.unitn.it}
\affiliation{Dipartimento di Fisica, Universit\`a di Trento
and CNR-INFM BEC Center, I-38050 Povo, Trento, Italy}
\author{S. D. Gensemer}
\affiliation{Van der Waals-Zeeman Instituut, Universiteit van
Amsterdam, Valckenierstraat 65, 1018 XE Amsterdam, The Netherlands}
\begin{abstract}
We propose an experimental method that allows for scaling of all the
diagonal correlation functions (such as the density and pair
correlation function) during the expansion of harmonically trapped
cold atomic gases with arbitrary interactions even if the trapping is
anisotropic. To do this we require that the interaction be tunable by
the use of a Feshbach resonance and that the trapping potential can be
manipulated independently along the axial and radial directions. This
permits a much greater accuracy and resolution in measurements of
correlation functions. We explain the explicit control procedure,
outline the scaling ansatz that we use and briefly discuss the systems
where it could be applied as well as possible experimental obstacles
to its implementation.

\end{abstract}

\maketitle 

How can we measure correlation functions in harmonically trapped cold
atomic gases? In order to measure them with good accuracy, the trapped
gas is first allowed to expand ballistically to improve optical
resolution and reduce its optical depth. However the interactions
present in the gas can in principle modify the correlation functions
significantly during the expansion and it is far from obvious how to
relate the measured values after expansion to those that existed in
the trap before the expansion \cite{jose}. Here we propose a new experimental
method whereby the release from the trap of an interacting gas is
accomplished in such a way that correlation functions exhibit
scaling during the expansion, and therefore their initial values can
be measured with great accuracy. We shall not discuss the off-diagonal
correlation functions such as the one-body density matrix $\langle
\psi^\dagger(\bfr) \psi(\bfr^\prime) \rangle$ with $\bfr \neq
\bfr^\prime$. However the present method is applicable to all real
space diagonal correlation functions.

In most experiments that measure the density $n(\bfr)$, the atomic gas
is allowed to expand and, when it is large enough for a satisfactory
optical resolution, it is imaged destructively. In many cases
(e.g. for ideal gases or interacting ones in the Thomas-Fermi limit),
the density after expansion is then related to the one before
expansion using {\em scaling} transformations
\cite{castindum,kagan}. By scaling we mean a coordinate transformation
of the kind:
\begin{equation}
x\to \lambda_x(t) x, \; y\to \lambda_y(t) y , \; z\to \lambda_z(t) z \label{eq:xyz}
\end{equation}
such that the density at a time $t$ after release from the trap can be
written in terms of the density at $t=0$:
\begin{equation}
\label{eq:n1scaling}
n({\bf r},t) = \alpha n_0(x/\lambda_x,y/\lambda_y,z/\lambda_z)
\end{equation}
The $\lambda_i(t)$ are known as scaling functions and
$\alpha=(\lambda_x \lambda_y \lambda_z)^{-1}$ is a time dependent
normalization factor.

While the density gives us very important information about such
things as thermodynamic properties and collective oscillations, it
does not reveal all the many-body aspects of the system. In fact we
can probe new physics by looking at other quantities such as the pair
correlation function which is defined as:
\begin{equation}
\gtwo({\bf r}_1,{\bf r}_2,t)=
  \frac{n^{(2)}({\bf r}_1,{\bf 
  r}_2,t)}{n({\bf r}_1,t)n({\bf r}_2,t)}
\end{equation}
with
\begin{equation}
n^{(2)}({\bf r}_1,{\bf
  r}_2,t)=\langle \psi^\dagger({\bf r}_1,t)
\psi^\dagger ({\bf r}_2,t)\psi({\bf
r}_2,t)\psi({\bf r}_1,t) \rangle
\end{equation}
where the $\psi$ are atomic field operators.  The pair correlation
function $g(\bfr_1,\bfr_2)$ is related to the probability of finding
an atom at $\bfr_2$ given that there is an atom at $\bfr_1$. If there
is more than one spin species present (as in the case of the
superfluid Fermi gas) then we can consider also the opposite spin pair
correlation function (see e.g. \cite{sandro}) where the atoms at
$\bfr_1$ and $\bfr_2$ are in different spin states. The pair
correlation function has been the subject of recent experimental
studies \cite{shimizu,arlt,aspect,esslinger,jin} while others have
focussed on correlations in momentum space \cite{bloch,jin}. The pair
correlation function can reveal new physics that is not measured by
the density. For example, it can reveal the ``bunching'' of thermal
bosons \cite{aspect}, a phenomenon which is also predicted to occur in
the unitary Fermi superfluid \cite{sandro}. The difficulty with
measuring $\gtwo(\bfr_1,\bfr_2)$ is that it has important features at
small values of $|\bfr_1-\bfr_2|$, sometimes comparable to the
interatomic distance. This requires very good optical resolution which
usually can only be achieved through an expansion of the gas.

This leads us naturally to the following question: as far as
real-space correlation functions are concerned (such as the density or
the pair correlation function), to what extent can they be faithfully
preserved during the expansion in such a way that a subsequent
measurement will be able to recover the values that existed in the
trap prior to the expansion?  For example, if we measure
$\gtwo(\bfr_1,\bfr_2,t)$ after an expansion time $t$, can we
reconstruct the initial $\gtwo(\bfr_1,\bfr_2,0)$ before the release
from the trap?

Ideally we would like to have a scaling relation like that of
Eq.(\ref{eq:n1scaling}) for $\gtwo$. Then we would have simply that
$\gtwo(\bfr_1,\bfr_2,t) \propto
\gtwo(r_{1i}/\lambda_i,r_{2i}/\lambda_i,0)$. Could such a scaling
relation exist? Unfortunately the answer is no in general, as shown in
\cite{sandro}: in an atomic gas, the existence of interactions imposes
certain boundary conditions on the many-body wave function and, as we
shall now show, these conditions are incompatible with scaling for all
correlation functions except the density. At distances shorter than
the interatomic separation, the wave function is determined up to a
proportionality constant by the Bethe-Peierls boundary condition
through which the interactions appear: for any value of the scattering
length $a$ the many-body wave function $\Psi$ obeys the condition for
all pairs of atoms $ij$:
\begin{equation}
\Psi(|{\bf r}_i-{\bf r}_j|\rightarrow 0) \propto
      \frac{1}{a}-\frac{1}{|{\bf r}_i-{\bf r}_j|} \label{eq:boundary}
\end{equation}
where the limit is taken keeping all other atoms and the center of
mass ${\bf r}_i+{\bf r}_j$ fixed. Having taken care of the
interactions via the boundary conditions, the only other condition on
$\Psi$ is that it must obey the free-particle Schr\"odinger
equation \cite{note}. These boundary conditions apply also to all diagonal
correlation functions: for example $\gtwo(\bfr_1,\bfr_2)$ obeys the
same condition as a function of $|\bfr_1-\bfr_2|$. The only exception
is the density $n(\bfr)$ since we need at least two coordinates to probe what
happens as two atoms come close to each other whereas the density
depends on a single coordinate. Eq.(\ref{eq:boundary}) has two
characteristic features: I) it is isotropic (since it depends on the
modulus of $\bfr_i-\bfr_j$ and not on its direction) and II) it introduces a
length scale $a$. These two features will pose problems for any
scaling solution of the form Eq.(\ref{eq:xyz}): I) requires that all
scaling functions $\lambda_i(t)$ be identical, which is in general
incompatible with the free expansion from an anisotropic trap. II)
means that, even if the trap is isotropic (leading to identical
$\lambda_i(t)$), a rescaling of the coordinates would change the ratio
$a/|{\bf r}_i-{\bf r}_j|$ which is required to be fixed. These
problems do not appear in the ideal gas case where $a=0$. If
$a\rightarrow \infty$ (e.g. the case of the unitary Fermi gas), II) is
of no consequence but I) still requires that the trap be spherical.

Nevertheless we can show that there exists in principle an
experimental procedure that can restore scaling during expansion to
the many-body wave function of a gas initially in equilibrium, trapped
in an {\it anisotropic} trap and interacting via an {\it arbitrary}
scattering length $a$. By this we mean that
$|\Psi(\bfr_1,...,\bfr_N,t)|^2 \propto
|\Psi(\bfr_1/\lambda(t),...,\bfr_N/\lambda(t),0)|^2$. Note that
information about the initial off-diagonal properties is lost and so
we cannot for example access the momentum space correlation functions
in the same way using this method.

To defeat problem I) we notice that it is not
the initial {\em trap} anisotropy (i.e. the difference between the
$\omega_i$) which destroys scaling but rather the {\em expansion}
anisotropy (the difference between the $\lambda_i(t)$). A judicious time
dependent modulation of the trapping constants $\omega_i(t)$ during
the release process is
sufficient to restore isotropy (all directions scaling with the same
$\lambda(t)$) as we shall see below. Problem II) can
also be dealt with if the scattering length can be changed during the
expansion by
modifying the applied magnetic field - which requires
the existence of an accessible Feshbach resonance. The idea is to
change the scattering length $a$ as a function of time during the
expansion so as to preserve Eq.(\ref{eq:boundary}).

We now briefly discuss the calculations which underlie the scaling
approach. Indeed, they follow almost exactly those in
\cite{castinunitary}. The main differences with respect to that
Reference are: here we consider a general anisotropic trap (as opposed
to a spherical one), a different time-dependence of the trapping
frequencies $\omega_i(t)$ and an arbitrary initial scattering length,
instead of the unitary limit. Therefore we shall give here only an
outline referring the reader for any further details to the more
complete discussion in that Reference.

\begin{figure}[b]
\begin{center}
\includegraphics[width=8cm]{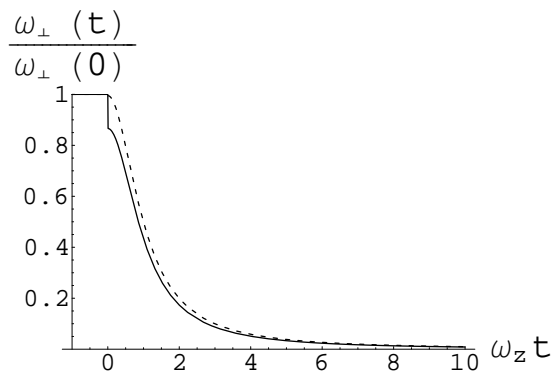}
\caption{Radial trapping frequency $\omega_\perp(t)$ from
  Eq.(\ref{eq:radial}) for $\omega_z(0)/\omega_\perp(0)=0.5$
  (continuous line) and $\omega_z(0)/\omega_\perp(0)=0.1$ (dashed
  line). Note that the $t=0$ discontinuity is appreciable in the first
  case (13\%) but negligible in the second (0.5\%).}
\label{fig1}
\end{center}
\end{figure}

For concreteness let us specialize to the experimentally relevant case
of an initially cigar-shaped potential (even though this assumption is
not necessary for our procedure since we could envision other
possibilities such as a pancake geometry):
\begin{equation}
U=\frac{1}{2}m\omega^2_\perp(t)\left(x^2+y^2\right)+\frac{1}{2}m\omega^2_z(t) z^2
\end{equation}
where for $t<0$ we have $\omega_x=\omega_y=\omega_\perp$ and $\omega_z
< \omega_\perp$. We also assume that, prior to expansion, the gas is
in a (many-body) eigenstate $\Psi(\bfr_1,...,\bfr_N,t=0)$ in
the trap $U(t=0)$, obeying the boundary conditions
Eq.(\ref{eq:boundary}) for a given scattering length $a$.
We make an ansatz for the time-dependence of the wave function:
\begin{eqnarray}
\Psi(\bfr_1,...,\bfr_N,t)=&&
 {\mathcal N}(t) \exp \left(i\sum_{j=1}^N mr^2_j \dot{\lambda}/2
\hbar \lambda\right) \nonumber\\
&&\times \Psi({\bf r}_1/\lambda,..., {\bf r}_N/\lambda,t=0)
\end{eqnarray}
where there is a time-dependent normalisation ${\mathcal N}$
and a single scaling factor $\lambda(t)$ which is the same for all
directions. For a general value of $\lambda(t)$ this wave function no
longer respects the boundary conditions for the initial value of the
scattering length and so we must change them using a Feshbach resonance
according to the expression
\begin{equation}
a(t)=a(0)\lambda(t) \label{eq:scatteringlength}
\end{equation}
Since $\lambda(t)$ generally increases with time, the scattering
length must also increase (in modulus) but the dilution parameter
$n|a|^3$ (where $n$ is the density of the gas) will remain
constant. This procedure solves problem II).  Now we insert this wave
function into the free-particle Schr\"odinger equation, use the
fact that $\Psi(t=0)$ is an eigenstate of the harmonic potential and,
by equating separately terms in $z^2 \Psi(t=0)$ and in $x^2 \Psi(t=0)$, we
find that it is indeed a solution provided that we can satisfy the two
equations
\begin{eqnarray}
\ddot{\lambda}&=&\frac{\omega_\perp^2(0)}{\lambda^3}-\omega_\perp^2(t) \lambda \nonumber\\
\ddot{\lambda}&=&\frac{\omega_z^2(0)}{\lambda^3}-\omega_z^2(t) \lambda \label{eq:scaling}
\end{eqnarray}
If a physically acceptable solution $\lambda(t)$ of these equations can be
found then also difficulty I) will have been overcome. Indeed such a
solution does exist {\it if the trapping constants
$\omega_{z,\perp}(t)$ can be changed independently in an appropriate
manner}. For example, one such solution is to simply switch off the
trapping potential along $z$ instantaneously as usual
($\omega_z(t>0)=0$) implying $\lambda(t)=\sqrt{1+\omega_z(0)^2t^2}$
(with $\lambda(0)=1$ and $\dot{\lambda}(0)=0$) which in
turn requires that
\begin{equation}
\omega_\perp(t)= \left\{ \begin{array}{ll}
\omega_\perp(0) & t<0 \\ & \\
\frac{\sqrt{\omega^2_\perp(0)-\omega_z^2(0)}}{1+\omega_z^2(0)t^2} & t>0 
\end{array} \right. \label{eq:radial}
\end{equation}
i.e., an initial discontinuous change of $\omega_\perp$ followed by a
gradual decrease to zero going as $1/t^2$ at large times. It is clear
that, for highly anisotropic traps, the discontinuity can be neglected
(see Fig.~\ref{fig1}).

So, to summarize, the experimental procedure consists of
simultaneously changing during the expansion the radial trapping
potential according to Eq.(\ref{eq:radial}) and the scattering length
according to Eq.(\ref{eq:scatteringlength}), i.e. $a(t)=a(0)
\sqrt{1+\omega_z(0)^2t^2}$.

We see therefore that the restriction due to the boundary condition is
not on the initial anisotropy of the trap but on the anisotropy of the
expansion. Here it proceeds isotropically for all times and
$|\Psi(t)|^2$ expands in a self-similar way so that the expansion acts
as a perfect microscope, enlarging without distortion all the features
of the cloud. We propose therefore that this method, when applicable,
be considered the ``best practice'' for measuring real-space
correlation functions.

To illustrate a practical implementation,
we consider the example of a Fermi gas of atoms, in a mixture of two
different spin states.  We consider an optical trap, in which the
atoms are trapped in the radial dimensions by a collimated laser beam,
and trapped in the axial dimension by another laser beam with a much
larger waist.  This combined optical potential would make independent
adjustment of the trapping frequencies relatively simple.  In
addition, a pair of magnetic coils in Helmholtz configuration would
provide the bias field needed to sweep the scattering length during
the expansion without contributing to the trapping potential.

By abruptly switching off the axial trapping potential, the cloud
would begin to expand freely along the axial dimension.  The radial
confinement would then be reduced by modulating the optical trap with
a pair of crossed acousto-optic modulators, which can be used to
produce a shallower, time-averaged potential which provides the
desired trapping frequency.  With this technique, we would expand the
cloud until its optical depth is low enough, and its radial dimensions
large enough, that atom noise correlations become measureable.  At
this point, the atoms would be imaged in situ using absorption
imaging.  By appropriately choosing the optical transition and
polarization for the imaging beam, atoms in the different spin states
could also be distinguished and imaged independently in rapid
succession.  In this way, spatial correlations between atoms in the
different spin states would be detected.

We now make a few final remarks. The possible candidate systems for
this method would include the dilute single species Bose gas and the
two spin or mass species Fermi gas at any value of the scattering
length. In this latter case it is not necessary that the the numbers
of atoms in each spin state be equal and so it could be also applied
to the polarized Fermi gas. On the other hand it would be perhaps
difficult to apply it to mixtures where more than one scattering
length is present since we might lose the ability to control each of
them simultaneously.

We relie of course on the existence of an accessible Feshbach
resonance and the application of a controlling magnetic
field. Therefore, if the trapping potential is magnetic in origin then
the two effects (trapping and detuning from resonance) must be handled
together. Finally we must keep in mind that the true potential seen by
the gas includes the effect of gravity. When the gas begins to expand
it will also begin to fall. This could introduce a distortion in the
potential seen by the gas since it would move from its original
position in the trap. To avoid this problem one possible solution
would be to introduce a compensating magnetic field gradient.

In conclusion, we began by discussing the importance and difficulty of
measurements of correlation functions at short distances which can
usually only be made after expansion. We then show that, apart from
the cases of the ideal and isotropic unitary gases, the second and
higher order diagonal correlation functions do not exhibit scaling
behaviour during the expansion of the gas and so scaling cannot be
used to relate the measured values after expansion to those existing
in the trap. We directly trace this violation of scaling to the
existence of the Bethe-Peierls boundary conditions imposed on the
many-body wave function due to the s-wave scattering between atoms. To
address this problem we propose an experimental procedure that fully
restores scaling to the diagonal correlation functions for systems
originally in anisotropic traps and interacting via an arbitrary value
of the scattering length which relies on a well-known ansatz for the
many-body wave function. This ansatz is valid provided that the
trapping potentials along different directions can be changed
independently and that the scattering length characterizing the
interaction can be changed as a function of time with the use of a
Feshbach resonance.

We acknowledge useful discussions with D. Blume, I. Carusotto, S. Giorgini,
A. Recati, S. Stringari and F. Werner. C. L. acknowledges support from
the Ministero dell'Istruzione, dell'Universit\`a e della Ricerca
(MIUR).


\begin{thebibliography}{99}
\bibitem{note} This description of the interacting atomic cold gas is, strictly
  speaking, only correct for the two species Fermi gas with s-wave
  interactions. In general, the many-body state may depend on other
  parameters beyond the scattering length. Nevertheless, it is
  believed that the contact boundary conditions provide a very good
  description in most other cases such as the weakly interacting Bose
  gas.
\end{thebibliography}
\end{document}